\documentclass[12pt]{article}
\usepackage{amsmath}
\usepackage{amssymb}
\usepackage{mathtext}
\usepackage{graphicx}

\begin{document}
\section*{KOVALEVSKAYA EXPONENTS AND POISSON STRUCTURES}\footnote{REGULAR
AND CHAOTIC DYNAMICS, V. 4, No.~3, 1999\\
{\it Received September 30, 1999\\
AMS MSC 34G20, 34L40}}\bigskip

\begin{centering}
A.\,V.\,BORISOV, S.\,L.\,DUDOLADOV\medskip\\
Faculty of Mechanics and Mathematics,\\
Department of Theoretical Mechanics
Moscow State University\\
Vorob'ievy gory, 119899 Moscow, Russia\\
E-mail: borisov@uni.udm.ru, dsl@online.ru\\
\end{centering}

\begin{abstract}
We consider generalizations of pairing relations for
Kovalevskaya exponents in quasihomogeneous systems with quasihomogeneous
tensor invariants. The case of presence of a Poisson structure in the system
is investigated in more detail.
We give some examples which illustrate general theorems.
\end{abstract}

\section{Quasihomogeneous systems. Kovalevskaya exponents}
A system of~$n$ differential equations
\begin{equation}
\label{Kov1}
{\dot x^i=v^i(x^1,\,\ldots \,,x^n)\,,\quad i=1,\,\ldots \, ,n\,,}
\end{equation}
is called {\it quasihomogeneous}
with quasihomogeneity exponents
$g_1,\,\ldots \, ,g_n,$  if
\begin{equation}
\label{Kov2}
v^i({\alpha}^{g_i}x^1,\,\ldots \, ,{\alpha}^{g_n}x^n)={\alpha}^{g_i+1}v^i(x^1,\,\ldots \, ,x^n)
\end{equation}
for all values of {\bf x} and $\alpha>0.$ Thus, the equations (\ref{Kov1})
are invariant under substitution $x^i\mapsto {\alpha}^{g_i}x^i,$ $t\mapsto
\frac{t}{\alpha}$
\cite{Yoshida}.

{\bf Remark 1.} {\it
A more general definition of quasihomogeneity of degree $m$ is the invariance
of the system~(\ref{Kov1}) under transformation
$x^i\mapsto\alpha^{g_i}x^i,$ $t\to \frac{t}{\alpha^{m-1}}$~\cite{KozlovFurta}.
All further results hold for this case as well.}

An important example of the equations (\ref{Kov1}),~(\ref{Kov2}) is a system
with quasihomogeneous quadratic  right-hand sides; in this case
${g_1=\,\ldots \, =g_n=1}$. Motion equations of many important problems of
dynamics (Euler-Poisson equations, Kirchhoff equations, Euler-Poincar\'e equations on
Lie algebras, Toda lattices, etc.) are of the quasihomogeneous form.

Differentiating (\ref{Kov2}) with respect to $\alpha$ and setting $\alpha=1$,
we obtain the Euler formula for quasihomogeneous functions:
\begin{equation}
\label{Kov3}
{\sum_{k=1}^n g_kx^k \frac{\partial v^i}{{\partial x^k}}=
(g_i+1)v^i\,,\quad i=1,\,\ldots \, ,n\,.}
\end{equation}

The equations (\ref{Kov1}) possess partial solutions
\begin{equation}
\label{Kov4}
{x^i=C_i t^{-g_i}\,,\quad i=1,\,\ldots \, ,n\,,}
\end{equation}
where the complex constants  $C_1,\,\ldots \, ,C_n$ should satisfy the
algebraic system of equations
\begin{equation}
\label{Kov5}
{v^i (C_1,\,\ldots \, ,C_n)=-g_i C_i\,,\quad i=1,\,\ldots \, ,n\,.}
\end{equation}
Let us write variational equations for the partial solution (\ref{Kov5}) as
\begin{equation}
\label{Kov6}
{\dot y}^i=\sum_{k=1}^n\frac{\partial v^i}{ {\partial x^k}}
(C_1 t^{-g_1},\,\ldots \, ,C_n t^{-g_n})y^k\,.
\end{equation}
The linear system (\ref{Kov6}) possesses partial solutions of the form
$$
{y^1={\varphi}^1
t^{\rho-g_1}\,,\,\ldots \, , y^n={\varphi}^n t^{\rho-g_n}\,,}
$$
where $\rho$ is an eigenvalue and  $\varphi$ is an eigenvector of a matrix
 ${\bf K}=\|K_j^i\|,$ $K_j^i=\Bigl(\frac{\partial v^i}{\partial x^j}({\bf
C})+g_i \delta_j^i\Bigr)$, $\delta_j^i$~is the Kronecker symbol.
The matrix~${\bf K}$ is called {\it Kovalevskaya matrix}, its eigenvalues
are called {\it Kovalevskaya exponents}\index{Показатели!Ковалевской}
(see \cite{Yoshida}). One of the Kovalevskaya exponents always equals
 $-1$ \cite{Yoshida}.

If the general solution of the system (\ref{Kov1}) is expressed in terms of
single-valued (meromorphic) functions of complex time, then
Kovalevskaya exponents, except for $-1$, are integer
(nonnegative integer, respectively).

Relations between Kovalevskaya exponents are
pointed out in \cite{KozTenzor}. They occur due to the presence of an
invariant tensor field in the system (\ref{Kov1}).

Recall, that a tensor field ${\bf T}$ of $(p,q)$ type is called
{\it quasihomogeneous} of degree $m$ with quasihomogeneity exponents
 $g_1,\,\ldots \, ,g_n,$ if
\[
T^{i_1\,\ldots \, i_p}_{j_1\,\ldots \, j_q}({\alpha}^{g_1}x^1,\,\ldots \,,
{\alpha}^{g_n}x^n)=
{\alpha}^{m-g_{j_1}-\,\ldots \, -g_{j_q}+g_{i_1}+\,\ldots \, +g_{i_p}}
T^{i_1\,\ldots \, i_p}_{j_1\,\ldots \, j_q}(x^1,\,\ldots \,,x^n)\,.
\]
This tensor field is {\it invariant} for the system (\ref{Kov1}), if its
Lie derivative along the vectorfield {\bf v} equals zero.

\section{Hamilton equations}
Let us consider quasihomogeneous equations of the form:
\begin{equation}
\label{Kov7}
{\dot x^i=\sum_k J^{ik}\frac{\partial H}{{\partial x^k}}\,,\quad
i=1,\,\ldots \, ,n\,,}
\end{equation}
where ${\bf J}=\| J^{ik}\|$ is a constant skew-symmetric tensor of type
(2,\,0),  $H$ is  a quasihomogeneous function of degree $m+1$:
\begin{equation}
\label{Kov8} H({\alpha}^{g_1}x^1,\,\ldots \,
,{\alpha}^{g_n}x^n)={\alpha}^{m+1}H(x^1,\,\ldots \, ,x^n)\,.
\end{equation}
Checking the fulfillment of the condition (\ref{Kov2}) and using (\ref{Kov8}),
we obtain
$$
\sum_{k=1}^n J^{ik}{\alpha}^{m+1-g_k}\frac{\partial H(x)}{{\partial
x^k}}= {\alpha}^{g_i+1}\sum_{k=1}^n J^{ik}\frac{\partial H(x)}{
{\partial x^k}}\,.
$$
Let ${\boldsymbol \Gamma}=\mbox{diag}(g_1, \dots, g_n).$
Differentiating the latter identity with respect to~$\alpha$ and setting
$\alpha=1,$ we obtain the following conditions in matrix form, to which
the quasihomogeneity exponents should satisfy:
\begin{equation}
\label{Kov9}
{\bf J}{\boldsymbol \Gamma}+{\boldsymbol \Gamma} {\bf J}=m {\bf J}\,.
\end{equation}
Let us note, that the equations (\ref{Kov7}) are Hamilton equations with
the Hamiltonian $H$ in  (possibly) noncanonical variables.
If ${\bf J}$ is a symplectic matrix, the conditions (\ref{Kov9}) have a
simple form:
$$
g_k+g_{k+\frac{n}{2}}=m\,.
$$
It is shown in \cite{Lochak,KozTenzor}, that in the case of diagonalizable
Kovalevskaya matrix its exponents satisfy the analogous relations
$$
\rho_k+\rho_{k+\frac{n}{2}}=m\,,
$$
moreover, there are always $-1$ and $m+1$ among them.

The following statement extends the corresponding
results~\cite{Lochak,KozTenzor} in general case of
nondiagonalizable matrix ${\bf K}.$  Let us note  that
one should not speak about single-valuedness and
meromorphy of the general solution  in the
nondiagonalizable situation.

{\bf Theorem 1.} {\it
Let ${\bf J}$ be a nondegenerate skew-symmetric matrix
for the equations \eqref{Kov7}. Then the Kovalevskaya exponents
decompose into pairs, which satisfy the relations}
$$
{{\rho}_k+{\rho}_{k+\frac{n}{2}}=m\,,\quad k=1,\,\ldots \, ,\frac{n}{2}\,,}
$$
{\it and the structure of Jordan cells, associated with the exponents~${\rho}_*$
and $(m-{\rho}_*)$, is the same.}

{\it Proof.}

Let us present the Kovalevskaya matrix in the form
${\bf K}={\bf JB}+{\boldsymbol \Gamma}$, where
$$
{\bf B}=\left\|\frac{{\partial}^2 H}{
{\partial x^i\partial x^k}}({\bf C})\right\|
$$
is a symmetric matrix. Therefore, all conclusions of Theorem~1 follow from
the chain of equivalent statements:
$$
\mbox{det}\|{\bf K}-\rho {\bf E}\|=0 \Leftrightarrow \mbox{det}\|{\bf K}-\rho {\bf E}\|^T=0 \Leftrightarrow
\mbox{det}\|({\bf K}-\rho {\bf E})^T {\bf J}^{-1}\|=0 \Leftrightarrow
$$
$$
\mbox{det}\|(-{\bf BJ}+{\boldsymbol \Gamma}-\rho {\bf E}){\bf J}^{-1}\|=0  \Leftrightarrow
\mbox{det} \|{\bf J}^{-1} ({\bf JB}+{\boldsymbol \Gamma}+(h-1-\rho){\bf E})\|=0\,.
$$
In the last link of the chain we used the fact, that
it is possible to substitute ${\bf J}^{-1}$ for ${\bf J}$ in
(\ref{Kov9}).$\blacksquare$

Let us generalize the above arguments on the case when the matrix $J^{ik}$
is not certainly nondegenerate and constant. It corresponds to the
considerations of quasihomogeneous systems which admit the Poisson structure
of a more general form (Lie-Poisson structures, quadratic structures, etc.)
Let us preliminary prove the main theorem.

{\bf Theorem 2.}
{\it Let us assume that the equations \eqref{Kov1} admit  a quasihomogeneous
tensor invariant~${\bf T}$ of degree $m$ and type $(2,0).$
Then natural numbers form $1$ to $n$ can be grouped in a set
$(k_1,\,\ldots \, ,k_n)$ so that $\rho_1,\,\ldots \, ,\rho_n$ satisfy
at least} $r=\mbox{rank } {\bf T}({\bf C})$ {\it relations:}
$$
{{\rho}_i+{\rho}_{k_i}=-m\,,\quad i=1,\,\ldots \, ,n\,.}
$$

{\it Proof}

Using the expression of Lie derivative for an invariant tensor field, it is
easy to show (see \cite{KozTenzor} for details) that the tensors ${\bf T}$ and
 ${\bf K}$ are connected by the following relations:
$$
-mT^{ij}=K_s^i T^{sj}+T^{is}K_s^j\,,
$$
which we shall write in matrix form
 \begin{equation}
\label{Kov10}
-m{\bf T}={\bf K T}+{\bf T K}^T,
\end{equation}
where ${\bf T}=\|T^{ij}\|, {\bf K}=\|K_{.j}^{i.}\|$.

Let the matrix ${\bf A}$ be formed of column-vectors $e_1,\,\ldots \, ,e_n,$
which are Jordan vectors of  ${\bf K}$:
$$
{\bf KA}={\bf AK}_{*}\,,
$$
where ${\bf K}_{*}$ for definiteness has the following form:
there are the Kovalevskaya exponents $(\rho_1,\,\ldots \, ,\rho_n)$
on the principal diagonal, and there can be units above the principal diagonal.
The analogous relation holds for the transposed matrix ${\bf K}^T$:
$$
{\bf K}^T({\bf A}^{-1})^T=({\bf A}^{-1})^T{\bf K}_{*}^T.
$$
Let us denote column-vectors, which form the matrix ${({\bf A}^{-1})}^T$, by
$f_1,\,\ldots \, ,f_n.$  On account of (\ref{Kov10}), we obtain
$$
 {\bf KT}({\bf A}^{-1})^T=-m{\bf T}({\bf A}^{-1})^T-{\bf TK}^T({\bf
 A}^{-1})^T= {\bf T}({\bf A}^{-1})^{T} (-m{\bf E}-({\bf K}_{*})^T)\,.
$$

The matrix  $(-m{\bf E}-({\bf K}_{*})^T)$ is also of Jordan form,
but now there can be $-1$ under the principal diagonal. Thus,
under transformation
${\bf A} \longmapsto {\bf T} {({\bf A}^{-1})}^T$
to Jordan vectors $(e_1,\,\ldots \, ,e_n)$ there correspond
 $r=\mbox{rank }{\bf T}({\bf C})$  independent vectors
$({\bf T}f_1,\,\ldots \, ,{\bf T}f_n),$ which are also Jordan vectors of
${\bf K}$ with the eigenvalues $(-m-{\rho}_1),\,\ldots \,
,(-m-{\rho}_n)$. $\blacksquare$

{\bf Remark 2.} {\it
It is  possible that $i = k_i$ in general case. Then
$\rho_i=-{\smash{\frac{m}{2}}}$. The following Corollary specifies the theorem in the case
of skew-symmetric tensor invariant.}

{\bf Corollary.}
{\it Let the tensor ${\bf T}$ be skew-symmetric. Then among natural numbers
from 1 to $n$ one can extract two subsets with distinct numbers}
 $(i_1,\,\ldots \, ,i_l)$ and $(k_1,\,\ldots \, ,k_l),$ $l=\frac{1}{2}\mbox{rank }{\bf
 T}({\bf C}),$ {\it such that Kovalevskaya exponents satisfy  $l$ relations}
 $$
 {\rho_{i_s}+\rho_{k_s}=-m\,,\quad s=1,\,\ldots \, ,l\,.}
 $$

Indeed, a nonzero vector ${\bf T} f_i$ can not be proportional to $e_i$
in the case of skew-symmetric ${\bf T}$. It follows from the fact that
 $(e_i, f_j)=\delta_{ij},$ where $(\cdot,\cdot)$ is the standard scalar product
in~$\mathbb{R}^n$ $({\bf A} {\bf A}^{-1}={\bf E}),$ and the skew-symmetric property of
 ${\bf T}$ implies $({\bf T}f_i,f_i)=0.$

Since the skew-symmetric structural tensor $J^{ij}$ is a tensor invariant
of motion equations for general Hamiltonian systems  (Section 2), then it follows from
the corollary that the Kovalevskaya exponents are coupled, and the number of
pairs equals $\frac{1}{2} \mbox{rank } {\bf J} ({\bf C}).$

\section{Invariant measure}
As a rule, quasihomogeneous equations of dynamics (the Euler-Poisson equations,
the Kirchhoff equations, etc.) possess an invariant measure besides
the degenerate Poisson structure (determined by algebra $e(3)$).
The existence of invariant measure imposes an additional condition on
Kovalevskaya exponents.

Indeed, let us assume that the system (\ref{Kov1}) admits a quasihomogeneous
tensor invariant of type~\mbox{$(n,0)$}
$$
{\Omega=\Omega ({\bf x}) dx^1\wedge \,\ldots \,  \wedge dx^n\,,\quad
\Omega ({\bf C})\ne 0\,.}
$$
Then ${\sum_{i=1}^n} \rho_i=m,$ where $m$ is the quasihomogeneity degree
of $\Omega$. If $\Omega$ is the standard measure, then
$\sum_{i=1}^n \rho_i=\sum_{i=1}^n g_i$; in particular, the sum of
Kovalevskaya exponents equals the system dimension $n$ for systems with
homogeneous quadratic right-hand sides. This result follows from the main
theorem of~\cite{KozTenzor}.

As it was pointed out in \cite{Kozlov16}, in the homogeneous case $(g_i=1)$
Kovalevskaya exponents are connected with multipliers of periodic solutions,
and their pairing for Hamiltonian systems follows from the
Poincar\'e-Lyapunov theorem on recurrence of roots of characteristic polynomial
of variational equations.

\section{Examples}

a) Let us consider a variant of {\it the system of Lotka-Volterra
type}~\cite{Volterra,Plank}, which can be written as
\begin{equation}
\label{Kov11}
\begin{array}{c}
{\dot x_i=x_i (\alpha_{i+1}x_{i+1}-\beta_{i-1}x_{i-1})\,,\qquad i=1,\,\ldots \,
,n,}\\[5pt]
{x_0=x_n\,,\qquad x_{n+1}=x_1\,,}
\end{array}
\end{equation}
where $\alpha_i, \beta_i$ are constants. The equations (\ref{Kov11}) are
generalizations of the integrable periodic Volterra system, for
which~${\alpha_i=\beta_i=const}$ \cite{Bogoyav}.

A straightforward calculation of Kovalevskaya exponents for the
system~(\ref{Kov11}) shows, that they satisfy the pairing conditions
$\rho_i+\rho_j=0$; it corresponds to occurrence of a quadratic tensor
invariant~$T^{ij}$ (by definition of quasihomogeneity degree).
However, the fulfillment of the theorem condition is not sufficient for
the presence of the Poisson structure. If the relation
$\prod_{i=1}^n\alpha_i=\prod_{i=1}^n\beta_i$ holds, the equations~(\ref{Kov11})
possess an additional linear integral  $F=(l,x),$ $l\in{\mathbb
R}^n,$ and under condition~$\alpha_i=\beta_i$ the system (\ref{Kov11}) is
indeed Hamiltonian with the quadratic Poisson bracket
$J^{ij}=C^{ij} x_i x_j$ and linear Hamiltonian.

b)
Let us consider {\it a generalized Suslov problem} as another example.
It describes a rigid body motion around a fixed point with the nonholonomic
constraint~$\omega_3=0$. If the center of mass
is on the principal axis, along which~$\omega_3=0$, then
motion equations of the system have the form:
\begin{equation}\label{7.12}
\begin{aligned}
{}\span I_1 \dot \omega_1 = \varepsilon \gamma_2\,, \qquad
I_2 \dot \omega_2 = -\varepsilon \gamma_1\,,\\
{}\span \dot \gamma_1 = -\omega_2 \gamma_3\,,\qquad
\dot \gamma_2 = \omega_1 \gamma_3\,,\qquad
\dot \gamma_3 = \omega_2 \gamma_1  -\omega_1 \gamma_2\,,
\end{aligned}
\end{equation}
where $I_1$, $I_2$ are components of the inertia tensor, $\varepsilon$ is
the distance from the fixed point to the center of mass. Calculation of
Kovalevskaya exponents gives the following values:

1. \quad   $\rho_1=-1,\,\rho_2=2,\,\rho_3=4,\,\rho_{4,5}=\frac{1}{2} \Bigl(
3 \pm \sqrt{1+8 \frac{I_2}{I_1}}\Bigr)$,

2. \quad   $\rho_1=-1,\,\rho_2=2,\,\rho_3=4,\,\rho_{4,5}=\frac{1}{2} \Bigl(
3 \pm \sqrt{1+8 \frac{I_1}{I_2}}\Bigr)$.

Similar in structure, but more complicated expressions for
the Kovalevskaya exponents
$\rho_{4,5}$ can be obtained in general case, when the position of the center of
mass and the nonintegrable constraint in the body are not related anyhow~\cite{70}.
These Kovalevskaya exponents are coupled:
$\rho_1+\rho_3=\rho_4+\rho_5=3$. Therefore, it is natural to expect
the presence of a tensor invariant in the system
\eqref{7.12} and possibility of its representation in the Hamiltonian form
\eqref{Kov7} with some in general nonconstant structural
tensor~$J^{ik}$. It is valid indeed, if one chooses the geometrical integral
 $F=\frac{1}{2} (\gamma_1^2+\gamma_2^2+\gamma_3^2)$ as the Hamiltonian.

Let $x=(x^1,x^2,x^3,x^4,x^5)=(\omega_1,\omega_2,\gamma_1,\gamma_2,\gamma_3)$
and rewrite~\eqref{7.12} as
\[
\dot x^i=J^{ij}(x)\frac{\partial F}{\partial x^j}\,,
\]
where
\[
J=\|J^{ij}\|=
\left(
\begin{array}{ccccc}
0&0&0&\frac{\varepsilon}{I_1}&0\\
0&0&-\frac{\varepsilon}{I_2}&0&0\\
0&\frac{\varepsilon}{I_2}&0&0&-\omega_2\\
-\frac{\varepsilon}{I_1}&0&0&0&\omega_1\\
0&0&\omega_2&-\omega_1&0\\
\end{array}
\right).
\]
For~\eqref{7.12} ${\bf J}$ is a tensor invariant of degree $3$,
which satisfies the
Jacobi identity, what one verifies by straightforward calculations.
The Casimir function of the Poisson structure~$J$ is the energy integral
(of the Suslov problem)
\begin{equation}\label{7.13}
\frac{1}{2}\bigl(I_1\omega_1^2+I_2\omega_2^2\bigr)+\varepsilon\gamma_2=h\,.
\end{equation}
The reduction on a symplectic sheet can be carried out explicitly,
if one notes, that the equations~\eqref{7.12} can be rewritten as Lagrange
equations \cite{Kozlov118} after excluding~$\gamma_3$ from~\eqref{7.13}
\[
\frac{d}{dt}\frac{\partial L}{\partial \dot \omega_i}=\frac{\partial L}{\partial \omega}\,,\qquad L=T-V\,,
\]
where
\[
T=\frac{1}{2} \bigl(
I_1^2\dot \omega_1^2+I_2^2\dot \omega_2^2\bigr)\,.
\]
These equations are not integrable for $I_1\ne I_2$. The Hamiltonian property
for the Suslov problem is rather unexpectable, since motion equations of
nonholonomic dynamics do not preserve the invariant measure in general case
\cite{KozlovVoz}.


c) Let us consider {\it a restricted problem in rigid body dynamics}.
\begin{equation}
\label{Dir16}
\begin{aligned}
\dot\omega_1=\omega_2\omega_3+z\gamma_2\,,\qquad
\dot\omega_2&=-\omega_1\omega_3-z\gamma_1\,,\qquad
\dot\omega_3&=-\gamma_2\,,\qquad
\dot\gamma&=\gamma\times\omega\,.
\end{aligned}
\end{equation}
The equations (\ref{Dir16}) at~$z=0$ were studied in \cite{KozTresh3}, where
their nonintegrability was shown with the help of the separatrix splitting
method; pictures of stochastic behavior are presented in \cite{BE}.

In general case when~$z\ne 0,$ the system (\ref{Dir16}) is quasihomogeneous
and possesses the following sets of partial solutions
$\omega_i=\frac{\omega_i^0}{t},$~$\gamma_i=\frac{\gamma_i^0}{t^2}$,    where
$$
\omega_1^0=0\,,\qquad\omega_2^0=0\,,\qquad\omega_3^0=2i\,,\\
\gamma_1^0=2\,,\qquad\gamma_2^0=2i\,,\qquad\gamma_3^0=0\,;
$$

and
$$
\omega_1^0=0\,,\qquad\omega_2^0=2i\,,\qquad\omega_3^0=0\,,\\
\gamma_1^0=\frac{2i}{z}\,,\qquad\gamma_2^0=0\,,\qquad\gamma_3^0=\frac{2}{z}\,.
$$
The Kovalevskaya exponents, corresponding to the chosen partial solutions, are
of the form:
$$
\rho=(-1, 0, 1, \rho_1, \rho_2, \rho_3)\,,
$$
where~$(\rho_1, \rho_2, \rho_3)$ are roots of the cubic equation
$\rho^3-9\rho^2+26\rho-(24+8z)=0,$ the solutions of which for
any~$z$ have a complicated algebraic form. Apparently, it  prevents from
the existence of algebraic integrals of motion for the system~(\ref{Dir16}).
The occurrence of a Poisson structure for the equations~\eqref{Dir16} has not
been also investigated.

d)
{Motion of a ferromagnet with the Barnett-London effect}
The essence of the quantum-mechanic effect of Barnett is that a neutral
ferromagnet magnetizes along the axis of rotation. In this case the magnetic
moment~${\bf B}$ is connected with its angular velocity~$\boldsymbol\omega$ by
relation~${\bf B}=\boldsymbol\Lambda_1\boldsymbol\omega,$
where~$\boldsymbol\Lambda_1$ is a symmetric linear operator. The analogous moment
occurs under rotation of a superconducting rigid body under the London effect.
If a body rotates in homogeneous magnetic field with the intensity~${\bf H}$,
then it is under magnetic forces with  the moment~${\bf B}\times {\bf H}.$
Let us denote
${\boldsymbol\gamma}={\bf H},$ then equations of motion can be written as:
\begin{equation}
\label{Id10}
\begin{array}{c}
\dot      {\bf      M}={\bf      M}\times      {\bf      AM}+{\boldsymbol\Lambda
{\bf M}}\times{\boldsymbol\gamma}\,,\qquad
\dot {\boldsymbol\gamma}={\boldsymbol\gamma}\times {\bf AM}\,,\\[5pt]
{\boldsymbol\Lambda}={\boldsymbol\Lambda}_1 {\bf A}\,,\qquad
{\bf A}={\bf I}^{-1}=\mbox{diag }(a_1,a_2,a_3)\,.
\end{array}
\end{equation}
As it is shown in~\cite{KozlovKzad}, the equations are Hamiltonian at
${{\boldsymbol\Lambda}={\bf A}}$ (they are reduced to the Kirchhoff equations, i.\,e.
to equations on the algebra~$e(3)$), and also at~${\boldsymbol\Lambda}=\mbox{diag }(\lambda_1,\lambda_2,\lambda_3),$
${\bf A}={\bf E}.$ In the last case they are integrable and  reduced
to the Clebsh case on the algebra~$e(3)$ by a linear coordinate
transformation \cite{Veselova}.

The equations (\ref{Id10}) possess  two integrals~$F_1=({\bf M},\boldsymbol\gamma),$
$F_2=(\boldsymbol\gamma,\boldsymbol\gamma)$ and the standard invariant measure.
There is the lack of two integrals for their integrability in general case.

These integrals are~${F_3=({\bf M,M})}$,
$F_4=({\bf M,AM})$  at~$\boldsymbol\Lambda=0$.  At
${\boldsymbol\Lambda=\mbox{diag }(\lambda_1,\lambda_2,\lambda_3)}$ using the method of
separatrix splitting, one can show that for  ${a_1\ne a_2\ne a_3\ne
a_1}$ the existence conditions for at least one of additional motion integrals,
generated by~$F_3$ or $F_4$, have the form
\begin{equation}
\label{Id11}
{\sum_{\hookrightarrow} \frac{\lambda_2-\lambda_3}{ a_1}=0\,,\qquad
\sum_{\hookrightarrow}
a_1^{-1}[a_2\lambda_3-a_3\lambda_2+\lambda_1(a_2-a_3)]=0\,.}
\end{equation}

It is obvious from (\ref{Id11}), that one more integral can actually exist
at~$\boldsymbol\Lambda={\bf E}.$ It is the integral of moment~$F_3=({\bf M,M}).$
The system~(\ref{Id10}) is completely
integrable at~${a_1=a_2=a}$,~$\boldsymbol\Lambda={\bf E}$,
and its additional integral is
$$
F_4=aM_3+\gamma_3\,.
$$
The question concerning the Hamiltonian property of the equations~(\ref{Id10})
was arisen in~\cite{Kozlov72}, however it has not been solved yet.
As it is noted in~\cite{Barkin},
the matrix ${\boldsymbol\Lambda}$ should be diagonal
${\boldsymbol\Lambda}=\mbox{diag }(\Lambda_1, \Lambda_2,\Lambda_3)$
for the Hamiltonian property
in the case of  $a_1\ne a_2\ne a_3\ne a_1$.

Calculation of Kovalevskaya exponents at~$a_2=a_3=B, a_1=1$  for
the solution
$$
(c_1,\,\ldots \,,c_6)=\left(0,\,\frac{1}{B}\sqrt{\frac{\lambda_3}{\lambda_2-\lambda_3}},\,
\frac{1}{B}\sqrt{\frac{\lambda_2}{\lambda_3-\lambda_2}},\,
\frac{c_3}{\lambda_2 c_2},\,
-\frac{Bc_2}{\lambda_3},\,-\frac{Bc_3}{\lambda_2}\right)
$$
gives the set
$(-1,1,2,2,1+\sqrt{B^2-2B},1-\sqrt{B^2-2B})$,
and for the solution
$$
(c_1,\,\ldots \,,c_6)=\left(i,\,\sqrt{\frac{\lambda_1}{B(\lambda_2-\lambda_3)}},\,
\sqrt{\frac{\lambda_1}{B(\lambda_3-\lambda_2)}},\,
0,\,
-\frac{Bc_2}{\lambda_1},\,-\frac{Bc_3}{\lambda_1}\right)
$$
the set
$(-1,2,2,2,B,1-B)$.

The pairing condition does not hold in this case, what is typical
(in general situation of nondegeneracy of the structural tensor
at the point~$(c_1,\,\ldots \,,c_6)$) for Hamiltonian systems.
However, this observation can not be considered to be a strong evidence of
the absence of the Hamiltonian property for the system (\ref{Id11}).

\end{document}